\documentclass[a4paper,10pt]{article}

\title{TORSION, DILATON AND GAUGE COUPLINGS}
\author{C. Mukku\footnote{University Grants Commission (India), Professor level Research Scientist.} \\International Institute of Information Technology\\Gachibowli, Hyderabad, India.}

\begin{document}

\maketitle

\begin{abstract}
Non-Abelian gauge fields are traditionally not coupled to torsion due to violation of gauge invariance. However, it is possible to couple torsion to Yang-Mills fields while maintaining gauge invariance provided one accepts that the gauge couplings then become scalar fields. In the past this has been untenable from experimental constraints at the current epoch for the electromagnetic field at least. Recent researches on the "landscape" arising out of string theory provides for many scalar fields which eventually determine the various low energy parameters including gauge couplings in the universe. With this scenario, we argue that the very early universe provides a Riemann-Cartan geometry with non-zero  torsion coupling to gauge fields. The torsion is just the derivative of gauge coupling (scalar) fields. As a result, in the evolution of the Universe, when the scalar (moduli) fields determine the geometry of the universe to be Riemannian, torsion goes to zero, implying that the associated modulus (and hence the gauge coupling) has a constant value. An equivalent view is that the modulus fixes the gauge coupling at some constant value causing the torsion to vanish as a consequence. Of course, when torsion vanishes we recover Einstein's theory for further evolution of the universe.
\end{abstract}

\section{Introduction.}
Theories of gravity with torsion have been studied since their inception with Cartan\cite{cartan, utiyama, kibble, sciama}. These theories have been plagued by a problem of plenty in the sense that the torsion fields in simple generalizations of the Einstein-Hilbert action did not propagate and requiring minimal propagating torsion raised the number of possible Lagrangian terms to $\approx 200$! In the simplest theories, the so called Einstein-Cartan-Sciama-Kibble (ECSK) theories \cite{hehl}, torsion was absent in the vacuum and coupling to Dirac fields showed it to be related to (spin-) angular momentum of the matter fields. Gauge fields required a separate treatment since their coupling to the gravitation field through the minimal coupling procedure necessarily brought in the antisymmetric part of the connection (hence torsion), causing a violation of the gauge principle and charge conservation. The traditional approach has been to assume that the minimal coupling procedure for gauge fields is carried out using the Christoffel symbols while the other matter fields couple through the complete non-symmetric connection. This isolation of the gauge fields from torsion seems unnatural and ideally one would like to end the isolation and treat gauge fields on par with other matter fields vis-a-vis interaction with torsion.
That this is indeed possible through the minimal coupling procedure while maintaining gauge invariance was shown in \cite{mukku}, generalizing a suggestion of \cite{hrrs} for modifying the minimal coupling procedure for the electromagnetic field. It was shown that the modification suggested by \cite{hrrs} was equivalent to keeping the minimal coupling procedure while converting the gauge coupling constant into a scalar field on spacetime. The torsion field takes a rather simple form as the derivative of the scalar field as we will see in the next section. For the particular case of electrodynamics, this spacetime dependent coupling would imply that the electric charge is a function on spacetime. It is obvious that this would have experimental consequences and puts restrictions on the variation of the electric charge over spacetime and essentially rules out torsion at the present epoch \cite{ni}.
The recent discovery of the "landscape" \cite{susskind} in string theory breathes new life into the above suggestions and we argue that one of the moduli from the landscape is to be identified with this torsion field and hence with the gauge charge. So, when the scalar field fixes the geometry to be Riemannian, with zero torsion, it also fixes the gauge coupling of the unified gauge group after the string compactification.
Thus, in the evolution of the universe, prior to the fixing of the moduli parameters for a particular universe, torsion is non-zero and is also given by a modulus. At a suitable stage in its evolution, as matter starts dissociating from radiation, all the gauge charges get fixed. I.e; the moduli for the gauge charges assume constant values. For our scenario, this also fixes the geometry to be Einsteinian with torsion becoming zero. The analysis for the descent from the grand unification scale through the electroweak to the current unbroken $U(1)$ symmetry follows the standard arguments \cite{polchinski}.
\section{Gauge fields and the Einstein-Cartan theory.}
A Riemann-Cartan spacetime is a spacetime endowded with a Riemannian metric $g_{\mu\nu}$ and parallel transport is carried out through a non-symmetric connection $\Gamma^{\mu}{}_{\nu\rho}$. The connection is assumed to be a metric connection, i.e; $\nabla_{\mu}g_{\nu\rho}=0$ where the covariant derivative of the connection is simply $\nabla_{\mu}=\partial_{\mu}-\Gamma_{\mu}$. In Einstein's theory, the connection is symmetric, $\Gamma^{\mu}{}_{\nu\rho}=\Gamma^{\mu}{}_{\rho\nu}$. For non-symmetric connections, the antisymmetric part $\frac{1}{2}(\Gamma^{\mu}{}_{\nu\rho}-\Gamma^{\mu}{}_{\rho\nu})$ is the torsion, $T^{\mu}{}_{\nu\rho}$.  Since our purpose is not to discuss torsion theories of gravity but their interaction with gauge fields, we simply note that all matter couplings are carried out through the usual minimal coupling procedure:\begin{equation}
\partial_{\mu}\longrightarrow \partial_{\mu}-\Gamma_{\mu}.
\end{equation}
We also note that the Riemann-Cartan connection can be written as:
\begin{equation}
\Gamma^{\sigma}_{\mu\alpha}=\{\begin{array}{l l l}\sigma&&\\
&\mu&\alpha
\end{array}\}-(T^{\sigma}{}_{\cdot}{}_{\mu}{}_{\alpha}-T_{\mu}{}_{\cdot}{}^{\sigma}{}_{\alpha}-T_{\alpha}{}_{\cdot}{}^{\sigma}{}_{\mu})
\label{cartangamma}
\end{equation}where $\{\begin{array}{l l l}\sigma&&\\ &\mu&\alpha \end{array}\}$ is the symmetric Riemannian connection (Christoffel symbols).

A Yang-Mills field is represented as:
\begin{equation}
F^{a}_{\mu\nu}=\partial_{\mu}A^{a}_{\nu}-\partial_{\nu}A^{a}_{\mu}+gC^{a}_{bc}A^{b}_{\mu}A^{c}_{\nu}
\end{equation} with $C^{a}_{bc}$ being the structure constants of the gauge group
and minimal coupling to the Einstein-Cartan gravitational field leads to:
\begin{equation}
F^{a}_{\mu\nu}=\partial_{\mu}A^{a}_{\nu}-\partial_{\nu}A^{a}_{\mu}+gC^{a}_{bc}A^{b}_{\mu}A^{c}_{\nu} -2A^{a}_{\rho}T^{\rho}_{\mu\nu}.\label{torsionF}
\end{equation}
The problem of gauge covariance of $F^{a}_{\mu\nu}$ is now manifest. Remembering that $F^{a}_{\mu\nu}$ is the curvature of a connection, we calculate it by assuming that the gauge coupling is a function on spacetime:
\begin{equation}
F_{\mu\nu}=[\partial_{\mu}-ig(x)A_{\mu},\partial_{\nu}-ig(x)A_{\nu}]
\end{equation}
This yields\begin{equation}
F^{a}_{\mu\nu}=-ig(x)[\partial_{\mu}A^{a}_{\nu}-\partial_{\nu}A^{a}_{\mu}+g(x)C^{a}_{bc}A^{b}_{\mu}A^{c}_{\nu}+(\partial_{\mu}ln(g(x))\delta^{\alpha}_{\nu}-\partial_{\nu}ln(g(x))\delta^{\alpha}_{\mu})].\label{torsiong}
\end{equation}
Comparision with equation(\ref{torsionF}) fixes the torsion field in terms of the coupling "function", $g(x)$:
\begin{equation}
T^{\alpha}_{\mu\nu}=\frac{1}{2}(\delta^{\alpha}_{\nu}\partial_{\mu}lng(x)-\delta^{\alpha}_{\mu}\partial_{\nu}lng(x)).
\end{equation}
We shall now call the landscape modulus that determines the gauge parameter, $\phi(x)$. The torsion tensor then reduces to the simpler form:
\begin{equation}
T^{\alpha}_{\mu\nu}=\frac{1}{2}(\delta^{\alpha}_{\nu}\partial_{\mu}\phi(x)-\delta^{\alpha}_{\mu}\partial_{\nu}\phi(x)) \label{torsionphi}
\end{equation}
and $g(x)=e^{\frac{\phi(x)}{2}}$.

Note that this form of the gauge coupling gives the traditional form for the dilaton coupling to gauge fields:
\begin{equation}
e^{-\phi(x)}F^{2}
\end{equation}

One may wonder if, in a unified framework, different coupling parameters would require different scalar (moduli) functions with the torsion being some suitable combinations of derivatives of the moduli.
That this is not the case can be seen easily through the example of the electroweak theory, with gauge group $SU(2)_{L}\times U(1)_{Y}$. Here, the covariant derivative can be written as:
\begin{equation}
D_{\mu}=\partial_{\mu}-igT^{a}A^{a}_{\mu}-i\frac{g'}{2}B_{\mu}
\end{equation}
without the coupling to gravity/torsion and the electric charge is given by\begin{equation}
e=\frac{g g'}{\sqrt{g^{2}+g'^{2}}}.
\end{equation}
If we now allow both $g$ and $g'$ to be functions on spacetime, then the curvature for this connection is simply given by:
\begin{eqnarray}
[D_{\mu},D_{\nu}]&=&-ig[\partial_{\mu}A_{\nu}-\partial_{\nu}A_{\mu}-ig[A_{\mu},A_{\nu}]-A_{\mu}\partial_{\nu}ln(g)+A_{\nu}\partial_{\mu} ln(g)]\\ \nonumber\label{ewcurvature}
&-&i\frac{g'}{2}[\partial_{\mu}B_{\nu}-\partial_{\nu}B_{\mu}-B_{\mu}\partial_{\nu}ln(g')+B_{\nu}\partial_{\mu} ln(g')].
\end{eqnarray}
Introducing gravity through the minimal coupling procedure modifies the right hand side of equation (12) to:
\begin{eqnarray}
&-&ig[\partial_{\mu}A_{\nu}-\partial_{\nu}A_{\mu}-ig[A_{\mu},A_{\nu}]-A_{\mu}\partial_{\nu}ln(g)+A_{\nu}\partial_{\mu} ln(g)-2A_{\rho}T^{\rho}_{\mu\nu}]\\ \nonumber\label{ewcurvature}
&-&i\frac{g'}{2}[\partial_{\mu}B_{\nu}-\partial_{\nu}B_{\mu}-B_{\mu}\partial_{\nu}ln(g')+B_{\nu}\partial_{\mu} ln(g')-2B_{\rho}T^{\rho}_{\mu\nu}].
\end{eqnarray} To ensure gauge invariance and hence charge conservations, it is clear that \begin{equation}
T^{\rho}_{\mu\nu}=\frac{1}{2}[\delta^{\rho}_{\nu}\partial_{\mu}ln(g)-\delta^{\rho}_{\mu}\partial_{\nu}ln(g)]
\end{equation} AND,\begin{equation}
T^{\rho}_{\mu\nu}=\frac{1}{2}[\delta^{\rho}_{\nu}\partial_{\mu}ln(g')-\delta^{\rho}_{\mu}\partial_{\nu}ln(g')].
\end{equation} These two equations can only be reconciled through the assumption that only one scalar field determines the gauge coupling and at this stage of the evolution of the universe, the couplings $g$ and $g'$ are indeed related.
This is a wonderful thing since the three coupling parameters for
the strong, weak and electromagnetic forces become equal at the
grand unification scale. From our perspective, this is natural! The
trifurcation of the single coupling then follows usual arguments
independent of the geometry of spacetime \cite{polchinski}.

We now have all the dynamical fields in our Riemann-Cartan theory. The Lagrangian which determines their dynamics is just the Einstein-Hilbert Lagrangian with non-zero torsion along with the Yang-Mills Lagrangian. Making explicit the torsion from the Ricci scalar, gives us the dynamics for the torsion and hence the scalar field $\phi(x)$ which we can identify with the dilaton.

\section{The reduced action.}
The minimalist approach dictates that the Riemann-Cartan action be simply the Einstein-Hilbert form with the Ricci scalar being replaced by the scalar curvature of the Riemann-Cartan spacetime, the Ricci-Cartan scalar.
\begin{equation}
S=\frac{1}{\kappa}\int \sqrt{-g} R(\Gamma)
\end{equation}
with $\Gamma$ being given by (\ref{cartangamma}).
It is not difficult to evaluate the Ricci-Cartan scalar in terms of the Ricci scalar ($R(\{\})$) and the torsion tensor. The action then takes the form:
\begin{equation}
S=\frac{1}{\kappa}\int \sqrt{-g} [R(\{\})+ g^{\mu\nu}(K_{\nu\sigma}^{\cdot \cdot \nu} K_{\mu\lambda}^{\cdot\cdot \sigma} - K_{\mu\sigma}^{\cdot \cdot \nu} K_{\nu\lambda}^{\cdot\cdot \sigma})]
\end{equation}
The tensor $K_{\mu\sigma}^{\cdot \cdot \nu}$, is a special combination of torsion tensors and is called the contortion tensor \cite{hehl}. With our notations, it is given by:
\begin{equation}
K_{\mu\nu}^{\cdot \cdot \lambda}=-T_{\mu\nu}{}^{\cdot \cdot
\lambda}+T_{\nu\cdot}{}^{\lambda}{}_{\mu}-T^{\lambda}{}_{\cdot}{}_{\mu\nu}
\end{equation}
and \begin{equation}
\Gamma^{\sigma}_{\mu\alpha}=\{\begin{array}{l l l}\sigma&&\\ &\mu&\alpha \end{array}\}-K_{\mu\nu}^{\cdot\cdot\lambda}
\end{equation}
With the form of torsion given by (\ref{torsionphi}), it is easy to write down the reduced Riemann-Cartan action as:
\begin{equation}
S=\frac{1}{\kappa}\int \sqrt{-g} [R(\{\})-24\partial_{\mu}\phi \partial^{\mu}\phi]
\end{equation}
\section{The complete action functional.}
Putting together all of the results given above, the complete (low energy) action functional for the Riemann-Cartan gravity coupled to Yang-Mills fields is given by:
\begin{equation}
S=\frac{1}{\kappa}\int \sqrt{-g} \{R(\{\})-24\partial_{\mu}\phi \partial^{\mu}\phi\} -\int\sqrt{-g}\frac{e^{-\phi}}{4}TrF^{2}
\end{equation}
Thus far, we have not addressed the interaction of torsion with
other matter fields, in particular Dirac fields. In the ECSK theory
\cite{hehl}, having excluded gauge fields from interacting with
torsion, it is found that torsion couples to the canonical spin
angular momentum tensor for the Dirac field. Hence, only the totally
antisymmetric part of torsion couples to Dirac fields. The
Lagrangian for a single Dirac field is simply: \begin{equation}
L_{Dirac}(\Gamma)=L_{Dirac}(\{\})+e\tau^{\alpha\beta\gamma}K_{\alpha\beta\gamma},\label{dirac}
\end{equation}
where the first term on the rhs is just the usual Lagrangian for a
Dirac field in a (pseudo-) Riemannian spacetime and the second term
is the interaction term for the spin-angular momentum of the Dirac
field with the torsion, while $e$ denotes the determinant of the
local tetrad. Since the spin-angular momentum is defined by:
\begin{equation}
\tau^{\alpha\beta\gamma}=\frac{1}{4}\overline{\psi}\gamma^{[\alpha}\gamma^{\beta}\gamma^{\gamma]}\psi,\end{equation}
it is clear that $\tau^{\alpha\beta\gamma}$ is totally
antisymmetric (the square brackets denote antisymmetrization with respect to all the enclosed indices). In our case, the torsion field takes a special form
when forced to couple to gauge fields and it is easy to see that its
totally antisymmetric part is zero, i.e; the second term on the
right hand side of equation(\ref{dirac}) vanishes. Therefore, the
inclusion of Dirac fields would follow the standard procedure for
coupling to Einsteinian gravity without torsion. Notice that the
Dirac fields are not excluded from interacting with torsion. It is
just that the special form of torsion imposed upon us by gauge
invariance, does not provide for interaction terms between Dirac
fields and the torsion field. This implies that in this
Riemann-Cartan theory, while torsion is non-zero and the minimal
coupling procedure holds for all matter fields, the particular form
that torsion ($\phi$) takes precludes it from interacting with Dirac
fields except through the (minimal) gauge coupling. Therefore,
evolution of all matter fields including gravity will follow the
Einsteinian path after the modulus $\phi$ becomes a constant as
matter is created in the early universe.

\section*{Acknowledgements.}
I would like to thank the IIIT and its director for hosting me and providing my salary from August 2005. I would also like to thank my mother, Mrs. Lakshmi Mukku for funding my research work. I would like to thank Profs. Bindu Bambah and Naresh Dadhich for comments on an earlier draft. I would also like to thank Prof. Bindu Bambah for support and encouragement.

\end{document}